\documentclass[submission,copyright,creativecommons]{eptcs}
\providecommand{\event}{INFINITY 2010} 

\usepackage{amsmath, amssymb}

\usepackage[ruled,vlined]{algorithm2e}
\LinesNumbered

\usepackage{graphicx}

\usepackage{pgf}
\usepackage{tikz}

\usetikzlibrary{arrows}

\usetikzlibrary{circuits}

\definecolor{turquoise}{rgb}{0 0.41 0.41}
\definecolor{rouge}{rgb}{0.79 0.0 0.1}
\definecolor{vert}{rgb}{0.15 0.4 0.1}
\definecolor{mauve}{rgb}{0.6 0.4 0.8}
\definecolor{violet}{rgb}{0.58 0. 0.41}
\definecolor{orange}{rgb}{0.8 0.4 0.2}
\definecolor{bleu}{rgb}{0.39, 0.58, 0.93}
\definecolor{gris}{rgb}{0.6,0.6,0.6}
\definecolor{grisfonce}{rgb}{0.4,0.4,0.4}
\definecolor{cpale1}{rgb}{1, 0.3, 0.3}
\definecolor{cpale2}{rgb}{0.3, 1, 0.3}
\definecolor{cpale3}{rgb}{0.3, 0.3, 1}
\definecolor{cpale4}{rgb}{1, 0.3, 1}
\definecolor{cpale5}{rgb}{1, 1, 0.3}
\definecolor{cpale6}{rgb}{0.3, 1, 1}
\definecolor{cpale7}{rgb}{0.9, 0.6, 0.2}
\definecolor{cpale8}{rgb}{0.7, 0.4, 1}
\definecolor{cpale9}{rgb}{0.5, 1, 0.75}
\definecolor{cpale10}{rgb}{0.8, 0.7, 0.6}
\definecolor{cpale11}{rgb}{0.6, 0.7, 0.8}
\definecolor{cpale12}{rgb}{0.2, 0.5, 0.9}
\definecolor{cpale13}{rgb}{0.5, 0.9, 0.2}
\definecolor{cpale14}{rgb}{0.9, 0.2, 0.5}
\definecolor{cpale15}{rgb}{0.7, 0.7, 0.7}
\definecolor{cpale16}{rgb}{0.8, 0.8, 0.5}

\newcommand{\A}{\mathcal{A}}

\newcommand{\grandr}{{\mathbb R}}
\newcommand{\grandrplus}{\grandr_{\geq 0}}

\newcommand{\tiling}{\mathit{Tiling}}

\newcommand{\Fleche}[1]{\stackrel{#1}{\Rightarrow}}
\newcommand{\steps}[0]{ {\rightarrow} }

\newcommand{\true}{{\tt true}}

\newcommand{\IM}{\mathit{IM}}
\newcommand{\BC}{\mathit{BC}}

\newcommand{\apron}{\textsc{Apron}}
\newcommand{\gdot}{\textsc{dot}}
\newcommand{\graphviz}{Graphviz}
\newcommand{\hytech}{{\sc HyTech}}
\newcommand{\imitator}{\textsc{Imitator}}
\newcommand{\imitatordeux}{\textsc{Imitator}\,II}

\newcommand{\ocaml}{OCaml}
\newcommand{\phaver}{PHAVer}
\newcommand{\polka}{NewPolka}

\newcommand{\python}{Python}
\newcommand{\red}{RED}
\newcommand{\TReX}{\textsc{TReX}}
\newcommand{\uppaal}{\textsc{Uppaal}}

\newcommand{\paragraphe}[1]{\paragraph{#1.}}

\title{IMITATOR~II: A Tool for Solving \\ the Good Parameters Problem in Timed Automata}
\author{\'Etienne Andr\'e\thanks{This work is partially supported by the VALMEM Project of Agence Nationale de la Recherche, grant ANR-06-ARFU-005, and by Institut Farman (ENS Cachan).}
\institute{LSV, ENS de Cachan \& CNRS, Cachan, France}
\email{andre@lsv.ens-cachan.fr}
}
\def\titlerunning{IMITATOR~II: A Tool for Solving the Good Parameters Problem in Timed Automata}
\def\authorrunning{\'E. Andr\'e}
\begin{document}
\maketitle

\begin{abstract}
We present here \imitatordeux{}, a new version of \imitator{}, a tool implementing the ``inverse method'' for parametric timed automata:
given a reference valuation of the parameters, it synthesizes a constraint such that, for any valuation satisfying this constraint, the system behaves the same as under the reference valuation in terms of traces, i.e., alternating sequences of locations and actions.
\imitatordeux{} also implements the ``behavioral cartography algorithm'', allowing us to solve the following good parameters problem: find a set of valuations within a given bounded parametric domain for which the system behaves well.
We present new features and optimizations of the tool, and give results of applications to various examples of asynchronous circuits and communication protocols.
\end{abstract}

\section{Introduction}

Timed automata~\cite{ad94} are finite control automata equipped with {\em clocks}, which are real-valued variables which increase uniformly, and that are compared with \emph{timing delays}.
One can check the correctness of a system modeled by a timed automaton for one particular value for each delay (using model checkers such as, e.g., \uppaal{}~\cite{lpy97}), but this does not give any information for other values.
Actually, checking the correctness of the system for all the delays, even in a bounded interval, would require an infinite number of calls to the model checker, because those delays can have real (or rational) values.
It is therefore interesting to reason \emph{parametrically}, by considering that these delays are unknown constants, or \emph{parameters}, and try to synthesize a {\em constraint} (i.e., a conjunction of linear inequalities) on these parameters which will guarantee a correct behavior of the system.
Such automata are called \emph{parametric timed automata} (PTA)~\cite{ahv93}. 

\paragraphe{The Good Parameters Problem for Timed Automata}
We aim at solving the \emph{good parameters problem}, as defined in~\cite{fjk08} in the framework of linear hybrid automata~\cite{achh92}:
``Given a PTA~$\A$ and a rectangular parameter domain $V_0$, what is the largest set of parameter values within $V_0$ for which~$\A$ is safe?''

The parameter design problem for timed automata (and more generally, for linear hybrid automata)
was formulated 
in~\cite{hw96}, where a straightforward solution is given,
based on the generation of the whole parametric state space until a fixpoint is reached.
Unfortunately, in all but the most simple cases, this is is prohibitively expensive due, in particular, to the brute exploration of the whole parametric state space.

In~\cite{fjk08}, the authors propose an extension based on the {\em counterexample guided abstraction refinement} (CEGAR, \cite{cgjlv00}).
When finding a counterexample, the system obtains constraints on the parameters that {\em make} the counterexample infeasible.
When all the counterexamples have been eliminated, the resulting constraints describe a set of parameters for which the system is safe.


\paragraphe{Contributions}
The tool \imitatordeux{} presented in this paper is based on the {\em inverse method}~\cite{acef09}, which starts from a ``good instantiation'' $\pi_0$ of the parameters that one wants to generalize.
More precisely, \imitatordeux{} synthesizes a constraint $K_0$ on the parameters that corresponds to a dense set of valuations such that, for all instantiation $\pi$ of parameters in this set, the behavior of the timed automaton $\A$ is {\em (time-abstract) equivalent} to the behavior of~$\A$ under~$\pi_0$, in the sense that they have the same trace sets.
This is useful to relax timing bounds, and gives a criterion of robustness.

Moreover, \imitatordeux{} implements the \emph{behavioral cartography algorithm}~\cite{af10}, which synthesizes a constraint on the parameters (``tile'') by calling the inverse method on integers point located within a given bounded parameter real-valued domain (rectangle)~$V_0$.
This algorithm allows us to partition the parametric space into a subset of ``good'' tiles (which correspond to ``good behaviors'') and a subset of ``bad'' ones.
Often in practice, what is covered is not only the 
{\em integer} subspace of~$V_0$, but two major extensions:
first, not only the integer points but a major part of the dense set of {\em real-valued} points of~$V_0$ is covered by the tiles;
second, the tiles are often unbounded w.r.t. several dimensions (hence are infinite), and cover most of the parametric space beyond~$V_0$, thus giving a solution to the good parameters problem.

\imitatordeux{} is a new version of \imitator{}~\cite{and09}, a prototype written in \python{}~\cite{python-web} implementing the inverse method, and calling the model checker \hytech{}~\cite{hhw97}.
\imitatordeux{} has been entirely rewritten and is now a standalone tool, making use of the \apron{} library~\cite{jm09} and the Parma Polyhedra Library~\cite{bhz08}.
Compared to \imitator{}, the computation timings of \imitatordeux{} have dramatically decreased.
Moreover, \imitatordeux{} offers new features, such as the implementation of the cartography algorithm, the visualization of the trace sets of the constraints, and of the cartography (for 2 parameter dimensions).


\paragraphe{Related Tools}
\imitatordeux{} has been designed to implement the inverse method and the cartography algorithm and, as far as we know, it is the only tool implementing this kind of algorithms.
Although it is thus not possible to compare directly the computation times of \imitatordeux{} with other tools, it is interesting to mention the following tools allowing to perform related analyses of timed systems.

\hytech{}~\cite{hhw97} is the first model checker for analyzing parametric hybrid automata. 
It features an intuitive input syntax, and performs reachability analysis and operations on states sets.
Although \hytech{} has been used to verify interesting case studies, it can hardly verify even medium sized examples because 
of its arithmetics with limited precision leading to overflows, and
its static composition of the automata, preventing the composition of more than a dozen of automata.

The tool \phaver{}~\cite{frehse05}, designed by Goran~Frehse, highly improves the scalability compared to \hytech{}, and performs analyses on parametric hybrid systems using exact arithmetics with unlimited precision and convex polyhedra, using the \emph{Parma Polyhedra Library} (PPL)~\cite{bhz08}.
Moreover, \phaver{} offers various features such as automatic partitioning, graphical outputs, and forward/backward abstraction refinement.
Various case studies have been verified, in particular in the framework of analog circuits~\cite{fkr06}.

\uppaal{} is a powerful tool for model checking timed automata extended with several data types~\cite{lpy97}.
In particular, it verifies very efficiently timing properties such as reachability, safety or liveness properties on timed automata.
However, although an extension allowing to perform parametric model checking is mentioned in~\cite{abbd01}, the standard version of \uppaal{} does not allow the use of hybrid or parametric systems.

\TReX{}~\cite{abs01} is a model checker allowing to verify properties on parametric timed automata extended with integer counters and finite-domain variables.
\TReX{} features on-the-fly verification of safety properties, as well as parameter synthesis either using parametric reachability, or in order to satisfy properties.
Various representations are allowed and both forward and backward exploration algorithms can be used.

Finally, the \red{} library~\cite{wang06} features analysis of real-time systems using Clock-Restriction Diagrams, as well as parametric analysis of hybrid systems using Hybrid-Restriction Diagrams.

\paragraphe{Plan of the Paper}
We first recall the framework of Parametric Timed Automata,
the inverse method algorithm and
the behavioral cartography algorithm in Section~\ref{sec:preliminaries}.
We then introduce \imitatordeux{} in Section~\ref{sec:implementation} and give details on its internal structure and its various features.
We present in Section~\ref{sec:experiments} a range of case studies including hardware devices and unbounded communication protocols.
We give final remarks in Section~\ref{sec:conclusion}.

\section{Behavioral Cartography of Timed Automata} \label{sec:preliminaries}


\paragraphe{Parametric Timed Automata} \label{ss:pta}

We use in this paper the same formalism as in~\cite{af10}.
Throughout this paper, we assume a fixed set $X = \{x_1, \dots, x_{H} \}$ of \emph{clocks},
and a fixed set $P = \{p_1, \dots, p_{M} \}$  of \emph{parameters}.
Given a constraint $C$ on the clocks and the parameters, the expression $\exists X: C$ denotes the constraint on the parameters obtained from~$C$ after elimination of the clocks.

Parametric timed automata are an extension of timed automata~\cite{ad94} to the parametric case, allowing within guards and invariants the use of parameters in place of constants~\cite{ahv93}.
A {\em parametric timed automaton (PTA)} $\A$ is a $6$-tuple of the form
\mbox{$\A=(\Sigma, Q, q_{0}, K, I, \steps)$},
where
$\Sigma$ is a finite set of actions,
$Q$ is a finite set of locations,
$q_{0} \in Q$ is the initial location,
$K$ is a constraint on the parameters, 
$I$ is the invariant assigning to every $q\in Q$
a constraint $I(q)$ on the clocks and the parameters, and
$\steps$ is a step 
relation consisting in elements of the form $(q,g,a,\rho,q')$  
 where
$q,q'\in Q$, $a\in\Sigma$, $\rho\subseteq X$ is a set of clocks to be reset by the step, and
$g$ (the step guard) is a constraint on the clocks and the parameters.

In the sequel, we consider the PTA $\A = (\Sigma, Q, q_{0}, K, I, \steps)$.
We simply denote this PTA by $\A(K)$, in order to emphasize the fact that only $K$ will change in $\A$.
For every parameter valuation \(\pi = (\pi_1, \dots, \pi_{M})\),
\(\A[\pi]\) denotes the PTA
\(\A(K)\),
where $K$ is $\bigwedge_{i = 1}^M p_i = \pi_i$.
This corresponds to the PTA obtained from~$\A$ by substituting every occurrence of a parameter \(p_i\) by constant \(\pi_i\) in the guards and invariants.
We say that $p_i$ is \emph{instantiated} with $\pi_i$.
Note that 
$\A[\pi]$ is a standard timed automaton.

A {\em (symbolic) state} $s$ of $\A(K)$ is a couple $(q, C)$ where $q$ is a location, and $C$ a constraint on the clocks and the parameters.
The \emph{initial state} of $\A(K)$ is a state $s_{0}$ of the form $ (q_{0}, C_{0})$,
where $C_{0} = K \land I(q_0) \land \bigwedge_{i = 1}^{H-1} x_i = x_{i+1}$.
In the latter expression, $K$ is the initial constraint on the parameters, $I(q_0)$ is the invariant of the initial state, and the rest of the expression lets clocks evolve from the same initial value.
A {\em run}~$R$ of $\A(K)$ 
is an alternating sequence of states and actions of the form
$(q_0, C_0) \Fleche{a_0} (q_1, C_1) \Fleche {a_1} \cdots \Fleche{a_{m-1}} (q_m, C_m)$,
such that for all $i = 0, \dots, m-1$,
$a_i \in \Sigma$ and $(q_i, C_i) \Fleche{a_i} (q_{i+1}, C_{i+1})$ is a step of $\A(K)$.
%
The \emph{trace associated to $R$} is the alternating sequence of locations and actions
$q_0 \Fleche{a_0} \cdots \Fleche{a_{m-1}} q_m$.
The \emph{trace set of $\A(K)$} refers to the set of traces associated to the runs of $\A(K)$.

In the following, we are interested in verifying properties on traces sets.
For example,
a trace can be said to be ``good'' if it never contains any ``bad'' location of a given set,
or if a given action always occurs before another one (see~\cite{af10}).
Given such a property on traces, we say that a trace is \emph{good} if it satisfies the property, and \emph{bad} otherwise.
Likewise, we say that a trace set is \emph{good} if all its traces are good, and \emph{bad} otherwise.
Actually, the good behaviors that can be captured with trace sets are relevant to {\em linear-time properties}~\cite{bk08}, which can express properties more general than reachability properties.

\paragraphe{The Inverse Method} \label{ss:im}



We recall here the inverse method algorithm $\IM(\A, \pi_0)$, as defined in~\cite{acef09}, which synthesizes a constraint~$K_0$ on the parameters such that $\pi_0 \models K_0$, and for all $\pi \in K_0$, the trace sets of $\A[\pi]$ and $\A[\pi_0]$ are equal.
Starting with $K = \true$,
we iteratively compute a growing set of reachable states.
When a $\pi_0$-\emph{incompatible} state $(q, C)$ is encountered (i.e., when \mbox{$\pi_0 \not \models (\exists X: C)$}), $K$ is refined as follows:
a $\pi_0$-incompatible inequality $J$ (i.e., such that \mbox{$\pi_0 \not \models J$}) is selected within the projection of $C$ onto the parameters 
and~$\neg J$ is added to~$K$.
The procedure is then started again with this new $K$, and so on, until no new state is computed.
We finally return the intersection of the projection onto the parameters of all the constraints associated to the reachable states.

The output of $\IM$ is a \emph{behavioral tile} in the following sense:
	A constraint~$K$ is said to be a {\em behavioral tile} (or more simply a {\em tile}), if 
	for all $\pi_1, \pi_2 \in K$, the trace sets of $\A[\pi_1]\textrm{ and }\A[\pi_2]$ are equal.
Note that a tile corresponds to a convex and dense set of real-valued points.
%
Given a tile $K$, the trace set of $\A(K)$ will be referred to as ``the trace set of~$K$''.

\IncMargin{1em}
\begin{algorithm}[ht!]
\SetKwInOut{Input}{input}\SetKwInOut{Output}{output}

\Input{A PTA $\A$ of initial state $s_0$} 
\Input{Valuation $\pi_0$ of the parameters}

\Output{Constraint $K$ on the parameters}

\BlankLine

$ i \leftarrow 0\,; \ \  K \leftarrow \true \,; \ \ S \leftarrow \{ s_0 \} $

\While{$\true$}{
	\While{there are $\pi_0$-incompatible states in $S$}
	{
		Select a $\pi_0$-incompatible state $(q, C)$ of $S$ (i.e., s.t. $\pi_0 \not\models (\exists X: C)$) \;
		
		Select a $\pi_0$-incompatible inequality $J$ in $(\exists X : C)$ (i.e., s.t. $\pi_0 \not\models J$) \;

		$K \leftarrow K \wedge\neg J $ \;
		$S \leftarrow \bigcup^i_{j = 0} \mathit{Post}^j_{\A(K)}(\{ s_0 \})$ \; \label{line:post_recomputation}
	} 
	
	\lIf{$\mathit{Post}_{\A(K)}(S) \sqsubseteq S $ \label{algo:IM:fixpoint}}
		{\Return{$K \leftarrow \bigcap_{(q, C) \in S} (\exists X: C) $}}

	$i \leftarrow i+1\,; \ \ S \leftarrow S \cup \mathit{Post}_{\A(K)}(S)$
	\tcp*{$S = \bigcup^i_{j = 0} \mathit{Post}^j_{\A(K)}(\{ s_0 \})$}

} 

\caption{$\IM(\A, \pi_0)$}\label{algo:IM}
\end{algorithm}\DecMargin{1em}

The algorithm~$\IM$ is given in Algorithm~\ref{algo:IM}.
%
%
We define $\mathit{Post}_{\A(K)}^i(S)$ as the set of states reachable from $S$ in exactly $i$~steps,
and $\mathit{Post}_{\A(K)}^*(S)$ as the set of all states reachable from $S$ in $\A(K)$
(i.e., $\mathit{Post}_{\A(K)}^*(S)=\bigcup_{i\geq 0 }\mathit{Post}_{\A(K)}^i(S)$).
Given two sets of states $S$ and $S'$, we write $S \sqsubseteq S' $ iff
$\forall s \in S , \exists s' \in S' $ s.t. $s = s'$.

\paragraphe{The Behavioral Cartography Algorithm} \label{ss:bc}


By iterating the above inverse method $\IM{}$ over all the \emph{integer} points of a rectangle $V_0$ (of which there are a finite number), one is able to decompose (most of) the parametric space included into $V_0$ into behavioral tiles.
We recall in Algorithm~\ref{algo:BC} the behavioral cartography algorithm, as defined in~\cite{af10}.

\IncMargin{1em}
\begin{algorithm}[ht!]
\SetKwInOut{Input}{input}\SetKwInOut{Output}{output}

\Input{A PTA $\A$, a finite rectangle $V_0 \subseteq \grandrplus^M$}

\Output{$\tiling$: list of tiles (initially empty)}

\BlankLine

\Repeat{$\tiling$ contains all the integer points of $V_0$}
{
	select an integer point $\pi \in V_0$\; 
	\lIf{$\pi$ does not belong to any tile of $\tiling$}{
		add $\IM(\A, \pi)$ to $\tiling$\;
	}
}

\caption{Behavioral Cartography Algorithm $\BC(\A, V_0)$ }\label{algo:tiling}
\label{algo:BC}
\end{algorithm}\DecMargin{1em}


In practice, most of (the real-valued space of) $V_0$ is covered by $\tiling$ (see case studies in Section~\ref{sec:experiments}), although some ``holes'' (i.e., small zones containing no integer point) may sometimes remain uncovered by~$\tiling$.
Furthermore, the space covered by $\tiling$ often largely exceeds the limits of~$V_0$. 

According to a given property on traces one wants to check, it is possible to partition trace sets between good and bad, and thus to partition
the rectangle $V_0$ into a good subspace 
(union of good tiles) and a bad subspace (union of bad tiles). 

The main advantage of this algorithm is that the cartography does not depend on the property one wants to check.
Only the partition between good and bad tiles does. 
Moreover, the algorithm 
does not compute the set of all the reachable states;
on the contrary, each call to~$\IM$ quickly reduces the state space by removing the ``bad'' states.
This allows us to overcome the state space explosion problem, which often prevents other methods, such as the computation of the whole set of reachable states 
to terminate. 

\section{Implementation} \label{sec:implementation}

\paragraphe{Features}

The input syntax of \imitatordeux{} to describe the network of PTAs modeling the system is given in~\cite{imitator2_web}, and is very close to the \hytech{} syntax.
%
%
\imitatordeux{} implements the ability to perform a full reachability analysis (computation of the set of all the reachable states), the inverse method algorithm, and the behavioral cartography algorithm.

\begin{figure}[ht!]
\centering
	\includegraphics[width=0.50\textwidth]{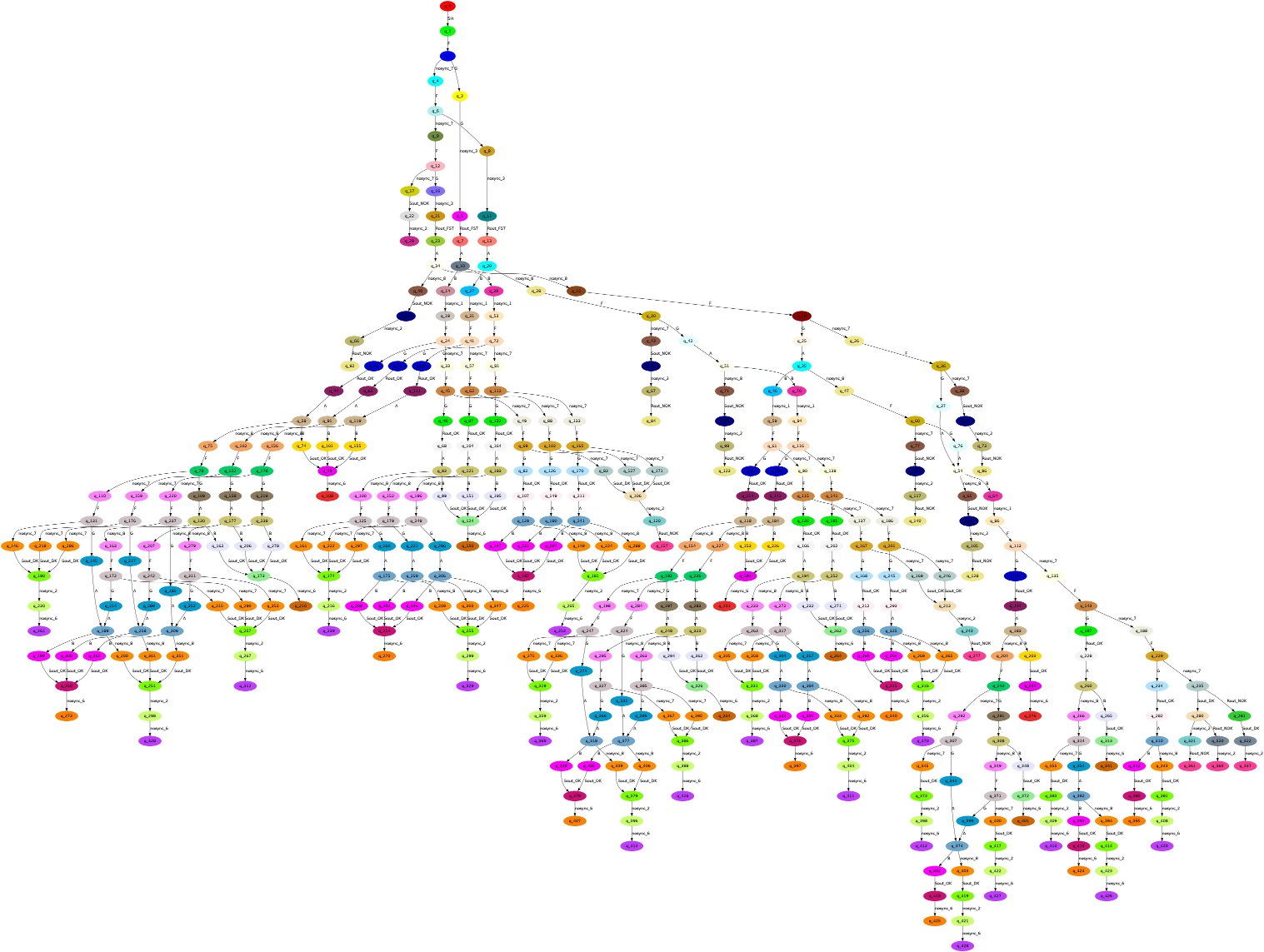}
	\ \ \ \
	\includegraphics[width=0.42\textwidth]{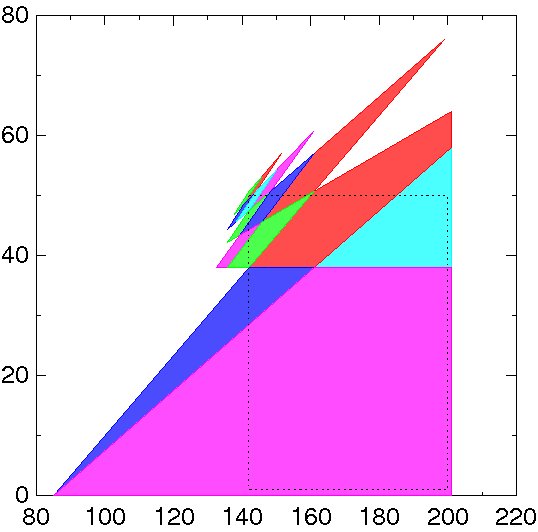}
\caption{Examples of trace set (left) and of cartography (right)}
\label{fig:graphical_outputs}
\end{figure}

When applying the inverse method, \imitatordeux{} takes as input a file describing the network of PTAs, and another file giving the reference valuation.
It synthesizes a constraint solving the inverse problem, as well as the corresponding trace set under a graphical form (see example in Figure~\ref{fig:graphical_outputs} left).
The description of all the parametric reachable states is also returned.

When applying the behavioral cartography, \imitatordeux{} takes as input a file describing the network of PTAs, and another file giving the reference rectangle, i.e., the bounds to consider for each parameter.
It synthesizes a list of tiles, as well as the trace set corresponding to each tile under a graphical form.
For systems with only two parameter dimensions, the cartography is also returned under a graphical form (see example in Figure~\ref{fig:graphical_outputs} right).
Two different modes can be considered for~$\BC$: (1) cover all the integer points of~$V_0$ or, (2) call a given number of times the inverse method on an integer point selected randomly within~$V_0$ (which is interesting for rectangles containing a very big number of integer points but few different tiles).
%
As shown in Table~\ref{table:features}, all those features (except the inverse method) are new features which were not available in \imitator{}.

\label{ss:features}

\begin{table}[ht!]
{

\centering
\small

\begin{tabular}{| c || c | c | c | c |}
	\hline
	Tool & Inverse Method & Cartography & Computation of traces & Graphical output \\
	\hline
	\imitator{} & yes & no & no & no \\
	\hline
	\imitatordeux{} & yes & yes & yes & yes \\
	\hline
\end{tabular}

}
\caption{Comparison of the features of \imitator{} and \imitatordeux{}}
\label{table:features}
\end{table}

Among the options for the tool (see~\cite{imitator2_web} for an exhaustive list), one can mention the possibility to add a limit for the depth of the $\mathit{Post}$ operation, or for the execution time, and an option for acyclic systems avoiding to check whether a state has been computed before.

\paragraphe{Implementation}

\imitatordeux{} is a tool written in \ocaml{},
making use of an external library for manipulating convex polyhedra, which can be,
depending on the user's preference, either the \polka{} library, available in the \apron{} library~\cite{jm09}, or the Parma Polyhedra Library (PPL)~\cite{bhz08}.
The trace sets, as well as the cartography for 2 parameter dimensions, are output under a graphical form using the \gdot{} module of the graph visualization software \graphviz{}~\cite{graphviz-web}.
\imitatordeux{} contains about 9000 lines of code, and its development took about 6 man-months.

States are represented using a triple $(q, v, C)$ made of the current location~$q$ in each automaton, a value for each discrete variable\footnote{Discrete variables are syntactic sugar allowing to factorize several locations into a single one. In~\imitatordeux{}, discrete variables are integer variables that can be updated using constants or other discrete variables.}~$v$, and a constraint $C$ on the clocks and the parameters.
In order to optimize the test of equality between a new computed state and the set of states computed previously, the states are stored in a hash table as follows:
to a given key $(q, v)$ of the hash table, we associate a list of constraints $C_1, \dots, C_n$, corresponding to the $n$ states $(q, v, C_1)$, \dots, $(q, v, C_n)$.
Contrarily to \hytech{},
\imitatordeux{} uses exact arithmetics with unlimited precision,
and performs an \emph{on-the-fly} composition of the automata,
allowing to analyze bigger systems, and decreasing drastically the computation time compared to \imitator{} (see Section~\ref{sec:experiments}).

\paragraphe{Optimization} \label{ss:optimizations}
Line~\ref{line:post_recomputation} in Algorithm~\ref{algo:IM} corresponds to the computation of all the states reachable in up to $i$~steps from~$s_0$, with the new constraint $K$ that has just been updated with the addition of some~$\neg J$.
However, this computation is redundant because no new state can be computed (because~$K$ has been restrained with~$\neg J$), and no state previously computed can be removed (because both $\neg J$ and the states previously computed are $\pi$-compatible).
Instead, we simply update the set~$S$ of states by adding~$\neg J$ to all the states computed, 
by replacing line~\ref{line:post_recomputation} in Algorithm~\ref{algo:IM} by the portion of algorithm given in Algorithm~\ref{algo:IM'}.

\IncMargin{1em}
\begin{algorithm}[ht!]

\ForEach{$(q, C) \in S$}
{
	$C \leftarrow C \land \neg J$
}

\caption{Modification of the Inverse Method Algorithm}\label{algo:IM'}
\end{algorithm}\DecMargin{1em}



\section{Case Studies} \label{sec:experiments}

We present in this section a range of case studies of asynchronous circuits and communication protocols.
The source code of \imitatordeux{} and various binaries, as well as the input file for all those case studies can be found in~\cite{imitator2_web}.
Experiments were conducted on an Intel Core2 Duo 2.4\,GHz with 2\,Gb.

\paragraphe{Inverse Method}
The results of the application of the inverse method to various case studies are given in Table~\ref{table:experiments-IM}.
We give from left to right the name of the example, 
the number of PTAs composing the global system $\A$,
the lower and upper bounds on the number of locations per PTA,
the number of clocks and parameters of $\A$,
of iterations of the algorithm,
of inequalities within~$K_0$,
of states and transitions,
the computation time in seconds using \imitator{},
and the computation time in seconds using \imitatordeux{}. 

\begin{table}[ht!]

\centering

\small

\begin{tabular}{|@{\,}c@{\,}||@{\,}c@{\,}|@{\,}c@{\,}|@{\,}c@{\,}|@{\,}c@{\,}||@{\,}c@{\,}|@{\,}c@{\,}|@{\,}c@{\,}|@{\,}c@{\,}||@{\,}c@{\,}|@{\,}c@{\,}|}
\hline
	Example & PTAs & loc./PTA &$|X|$ & $|P|$ & iter. & $|K_0|$ & states & trans. & Time1 & Time2 \\
\hline
	SR-latch & 3 & $[3, 8]$ & 3 & 3 & 5 & 2 & 4 & 3 & $0.11$ & $0.007$ \\
\hline
	Flip-flop \cite{cc07} & 5 & $[4, 16]$ & 5 & 12 & 9 & 6 & 11 & 10 & $1.6$ & $0.122$ \\
\hline
	And--Or \cite{cc05} & 3 & $[4, 8]$ & 4 & 12 & 14 & 4 & 13 & 13 & $1.81$ & $0.15$ \\
\hline
	Valmem Latch & 7 & $[2, 5]$ & 8 & 13 & 12 & 6 & 18 & 17 & $14.4$ & $0.345$ \\
\hline
	CSMA/CD \cite{knsw07} & 3 & $[3, 8]$ & 3 & 3 & 19 & 2 & 219 & 342 & $41$ & $1.01$ \\
\hline
	RCP \cite{kns03} & 5 & $[6, 11]$ & 6 & 5 & 20 & 2 & 327 & 518 & 64 & $2.3$ \\
\hline
	SPSMALL$_1$ \cite{cefx09} & 10 & $[3, 8]$ & 10 & 26 & 32 & 23 & 31 & 30 & 4680 & 2.6 \\
\hline
	BRP \cite{dkrt97} & 6 & $[2, 6]$ & 7 & 6 & 30 & 7 & 429 & 474 & 901 & 34 \\
\hline
	SPSMALL$_2$ \cite{cefx09} & 28 & $[2, 11]$ & 28 & 5 & 92 & 8 & 472 & 548 & - & 1755 \\
\hline
\end{tabular}

\caption{Summary of experiments for the inverse method}
\label{table:experiments-IM}
\end{table}

The SPSMALL case study corresponds to an asynchronous memory sold by ST-Microelectronics, and studied in the framework of VALMEM project.
We considered two versions of this case study: the first one (``SPSMALL$_1$'') was manually abstracted from the VHDL code (see~\cite{cefx09}) and several gates have been merged into a single PTA.
The second model (``SPSMALL$_2$'') has been automatically generated from the VHDL code without any simplification.
It is impossible to analyze SPSMALL$_2$ using the first version of \imitator{} because \hytech{} runs out of memory when trying to statically compose the 28~automata in parallel.

The Valmem latch is an example of latch studied in the framework of VALMEM project.

Note that the computation time using \imitatordeux{} has dramatically decreased compared to \imitator{} for all examples: the time has been divided at least by 10, and up to 2000 for the SPSMALL$_1$ memory.
Explanations for this high improvement are the rewriting of the tool using a library of convex polyhedra instead of the call to \hytech{}, the on-the-fly composition of the different PTAs, and the optimization of the algorithm described in Section~\ref{ss:optimizations}.

\paragraphe{Behavioral Cartography}
The results of the application of the behavioral cartography algorithm to various case studies are given in Table~\ref{table:experiments-BC}.
We give from left to right the name of the example,
the number of PTAs composing the global system $\A$,
the lower and upper bounds on the number of locations per PTA,
the number of clocks 
of~$\A$ (those first 4~columns are identical to Table~\ref{table:experiments-IM}),
the number of parameters varying in the cartography,
of integer points within $V_0$,
of tiles computed,
the average number per tile of states and transitions of the trace set,
and the computation time in seconds using \imitatordeux{} (since the cartography is a new feature available in \imitatordeux{} only, no comparison is possible with \imitator{}).

\begin{table}[ht!]

\centering

\small

\begin{tabular}{|@{\,}c@{\,}||@{\,}c@{\,}|@{\,}c@{\,}|@{\,}c@{\,}|@{\,}c@{\,}|@{\,}c@{\,}||@{\,}c@{\,}|@{\,}c@{\,} | @{\,}c@{\,} ||@{\,}c@{\,}|}
 \hline
   Example & PTAs & loc./PTA & $|X|$  & $|P|$ & $|V_0|$ & tiles & states & trans. & Time \\
 \hline
   SR-latch & 3 & $[3, 8]$ & 3 & 3 & $1331$ & 6 & 5 & 4 & $0.3$ \\
 \hline
   Flip-flop \cite{cc07} & 5 & $[4, 16]$ & 5 & 2 & $644$ & 8 & 15 & 14 & 3 \\
 \hline
   And--Or \cite{cc05} & 3 & $[4, 8]$ & 4 & 6 & $75600$ & 4 & 64 & 72 & 118 \\
 \hline
   Valmem Latch & 7 & $[2, 5]$ & 8 & 4 & $73062$ & 5 & 21 & 20 & $96.3$ \\
 \hline
   CSMA/CD \cite{knsw07} & 3 & $[3, 8]$ & 3 & 3 & $2000$ & 140 & 349 & 545 & 269 \\
 \hline
   RCP \cite{kns03} & 5 & $[6, 11]$ & 6 & 3 & $186050$ & 19 & 5688 & 9312 & 7018 \\
 \hline
   SPSMALL$_1$ \cite{cefx09} & 10 & $[3, 8]$ & 10 & 2 & $3149$ & $259$ & $60$ & $61$ & $1194$ \\
 \hline
\end{tabular}

\caption{Summary of experiments for the cartography algorithm}
\label{table:experiments-BC}
\end{table}

For all those examples, the cartography covers 100\,\% of the real-valued space of $V_0$, except for the Root Contention Protocol (see Section~\ref{ss:bc}), where ``only'' 99,99\,\% of $V_0$ is covered.
Moreover, a significant part of the real-valued space outside $V_0$ is also covered.

Note that it is possible to find examples for which the algorithm~$\BC$ does not terminate for some~$V_0$, because the algorithm~$\IM$ does not terminate for some~$\pi \in V_0$.
This is in particular the case of the ``And--Or'' circuit considered in~\cite{cc05}, for a different~$V_0$ from the one considered in Table~\ref{table:experiments-BC}.

\section{Conclusion} \label{sec:conclusion}

\imitatordeux{} allows us to solve the good parameters problem for timed automata by iterating the inverse method on the integer points of a real-valued parameter domain~$V_0$.
In practice, our cartography algorithm covers not only (most of) $V_0$ but also a significant part of the whole parametric space beyond~$V_0$.
The tool has been successfully applied to various examples of asynchronous circuits and protocols.

\paragraphe{Ongoing and Future Work}
Ulrich~K\"uhne is currently extending \imitatordeux{} to hybrid systems, where clocks evolve at different rates.
Romain~Soulat is currently implementing variants and optimizations of~$\IM$ in order to verify larger asynchronous memory circuits, in particular using an on-the-fly intersection of the constraints associated to the states, allowing to merge states.
A variant of~$\IM$ is also under implementation, where the fixpoint condition (line~\ref{algo:IM:fixpoint} of Algorithm~\ref{algo:IM}) is modified as follows: instead of checking whether all new states are equal to states computed previously, we check whether all new states are \emph{included} (in the sense of constraint inclusion) into former states.

Future work include the automatic partition into good and bad tiles, using an external tool such as \uppaal{}. 
%
We are also studying a ``dynamic'' cartography, where the space unit between the selected points (so far, one integer) can be refined in order to fill the remaining holes.
%
It would also be interesting to reason in a backward manner, i.e., considering a $\mathit{Pre}$ operation instead of $\mathit{Post}$ in Algorithm~\ref{algo:IM}.


\paragraphe{Acknowledgments}
I thank Bertrand~Jeannet for his help to link \imitatordeux{} with \apron{}, Ulrich K\"uhne for the interface with PPL, and Daphn\'e~Dussaud for the graphical output for the cartography.
I~also thank anonymous referees for their helpful comments.

\bibliographystyle{eptcs} 
\bibliography{biblio}

\end{document}